%% file: main.tex
\documentclass[10pt,twocolumn,letterpaper]{article}
\usepackage[pagenumbers]{cvpr}
\usepackage{placeins}
\input{preamble}

\title{STAC++: Plug-and-Play Adaptive Spatio-Temporal Aware Cache Compression for Streaming 3D Reconstruction}
\author{%
  Boyi Zhu\textsuperscript{1}\qquad
  Youcheng Cai\textsuperscript{1}\qquad
  Dong Wang\textsuperscript{2}\\[3pt]
  {\normalsize\textsuperscript{1}University of Science and Technology of China\qquad
   \textsuperscript{2}Anhui Jianzhu University}%
}
\hypersetup{
  pdftitle={STAC++: Plug-and-Play Adaptive Spatio-Temporal Aware Cache Compression for Streaming 3D Reconstruction},
  pdfauthor={Boyi Zhu, Youcheng Cai, Dong Wang}
}
\begin{document}
\maketitle
\input{sec/0_abstract}

\input{sec/1_intro}
\input{sec/2_related}
\input{sec/3_preliminaries}
\input{sec/4_method}
\input{sec/5_experiments}

\input{sec/6_conclusion}

\section*{LLM Usage Statement}
OpenAI Codex assisted with organizing the manuscript, editing language, checking correspondence between the description and the implementation, and preparing LaTeX sources. The authors are responsible for validating the method, experiments, citations, and final claims.

{\small
\bibliographystyle{ieeenat_fullname}
\bibliography{references}
}
\end{document}

%% file: preamble.tex
\definecolor{cvprblue}{rgb}{0.21,0.49,0.74}
\usepackage[pagebackref,breaklinks,colorlinks,allcolors=cvprblue]{hyperref}
\usepackage[capitalize]{cleveref}

\newcommand{\method}{STAC++\xspace}

%% file: sec/0_abstract.tex
\begin{abstract}
Streaming 3D reconstruction must preserve geometry under limited memory, yet STAC's fixed-resolution spatial key--value (KV) cache grows as scenes expand.
We propose \textbf{STAC++}, a plug-and-play, training-free framework that compresses and reuses spatial memory for long-term reconstruction in unbounded scenes.
\textbf{Adaptive Spatial KV Cache (ASKC)} organizes KV states into an octree with adaptive resolutions across regions and layers.
\textbf{Response-Aware Cache Compression (RACC)} guides KV merging by measuring attention-output changes, balancing storage savings with geometric preservation.
\textbf{Temporal Context Refinement (TCR)} reuses cached intermediate features and recomputes only a short backbone suffix to refine historical predictions with later observations.
Integrated into four pretrained streaming backbones, STAC++ reduces spatial memory consumption while maintaining competitive reconstruction quality on long sequences and large scenes, with further quality gains from refinement at limited computational overhead.
The code will be made publicly available upon acceptance.
\end{abstract}

%% file: sec/1_intro.tex
\section{Introduction}
\label{sec:intro}

3D reconstruction is a fundamental problem in computer vision and computer graphics, with a wide range of applications including robotic mapping~\cite{elasticfusion}, augmented reality~\cite{kinectfusion}, and novel-view synthesis~\cite{nerf,gaussian_splatting}. Traditional image-based 3D reconstruction combines Structure-from-Motion (SfM)~\cite{schoenberger2016sfm} for camera estimation and sparse structure recovery with Multi-View Stereo (MVS)~\cite{schoenberger2016mvs} for dense geometry. These pipelines rely on explicit correspondence matching and iterative geometric optimization. Recent feed-forward geometric models instead infer scene geometry directly from images using pretrained networks, reducing reliance on scene-specific iterative optimization and improving reconstruction efficiency.

Feed-forward reconstruction methods can be broadly categorized into offline methods that process a fixed image collection and online methods that process incoming images sequentially.
Offline methods include DUSt3R~\cite{dust3r}, which predicts pairwise pointmaps followed by global alignment, and VGGT~\cite{vggt}, which jointly predicts cameras and dense geometry from multiple views.
However, their computational and memory costs limit scalability to long sequences.
Online methods~\cite{stream3r,streamvggt,infinitevggt,lingbotmap} instead reuse historical representations to reconstruct incoming frames incrementally.
StreamVGGT~\cite{streamvggt} adapts VGGT to streaming inference through causal attention and cached historical key--value (KV) states, but its cache grows with sequence length.
InfiniteVGGT~\cite{infinitevggt} controls this growth with a bounded rolling cache and diversity-aware token pruning, enabling long-sequence inference within a fixed memory budget.
However, these methods do not explicitly exploit 3D spatial information in KV-cache management.
STAC~\cite{stac} introduces a spatial KV cache that merges historical tokens into voxel-aligned representatives, reducing spatial redundancy while preserving geometric evidence.
However, its fixed-resolution grid allows the number of occupied voxels to grow with scene extent, so long streams exploring unbounded environments can still exhaust GPU memory.
Uniform voxel coarsening saves memory but risks losing fine geometric detail in complex regions.

\input{fig/teaser}

STAC uses spatial memory to reconstruct incoming frames, yet earlier predictions remain fixed even when later views reveal the same surfaces more clearly.
These limitations motivate our central question: \emph{How can we adaptively compress and actively reuse spatial KV memory to support long-term reconstruction of unbounded scenes, while preserving geometric detail and refining earlier predictions under limited memory and computational budgets?}
Answering this question poses two challenges.
(1) Handling unbounded scenes requires spatial representations that accommodate continually expanding scene coverage while controlling cache growth and preserving geometric detail.
(2) Operating within a bounded memory budget requires retaining informative historical evidence and efficiently reusing it to refine earlier predictions with later observations, without replaying the entire backbone.

To address these challenges, we propose \textbf{STAC++}, a plug-and-play, training-free framework that controls spatial memory growth and efficiently reuses historical evidence for long-term streaming reconstruction in unbounded scenes (see Fig.~\ref{fig:octree_memory}).
Firstly, \textbf{Adaptive Spatial KV Cache (ASKC)} organizes spatial KV states into an octree hierarchy with different resolutions across scene regions and transformer layers.
This hierarchy supports merging fine-grained cells into coarser representations as the scene expands.
Secondly, \textbf{Response-Aware Cache Compression (RACC)} guides KV merging by evaluating how candidate merges change the backbone's attention outputs (hereafter referred to as \textbf{responses}).
It identifies spatial representations that can be consolidated with limited response changes, balancing memory savings against geometric information preservation.
Finally, \textbf{Temporal Context Refinement (TCR)} revisits historical frames after a fixed delay, using later observations and accumulated spatial memory to refine earlier predictions.
By reusing cached intermediate features and recomputing only a short backbone suffix, TCR creates additional opportunities to exploit spatial memory while limiting the computation required for refinement.
We implement STAC++ as an inference-time plugin for STream3R~\cite{stream3r}, StreamVGGT~\cite{streamvggt}, InfiniteVGGT~\cite{infinitevggt}, and LingBot-Map~\cite{lingbotmap}, preserving their pretrained weights without additional training. Experiments across these four backbones demonstrate that STAC++ reduces spatial memory consumption while maintaining competitive reconstruction quality, scales to long sequences and large scenes, and further improves reconstruction quality through delayed suffix refinement with limited computational overhead.

Our main contributions are summarized as follows:
\begin{itemize}
    \item We propose \textbf{STAC++}, a \textbf{plug-and-play}, training-free framework that unifies adaptive spatial memory compression and historical evidence reuse for long-term streaming reconstruction in unbounded scenes without modifying pretrained backbone weights.

    \item We develop \textbf{ASKC}, \textbf{RACC}, and \textbf{TCR} for hierarchical spatial KV caching, response-guided merging, and suffix refinement of historical predictions, controlling memory and computation while preserving geometry.
\end{itemize}

%% file: fig/teaser.tex
\begin{figure}[t]
    \centering
    \includegraphics[pagebox=cropbox,width=\columnwidth]{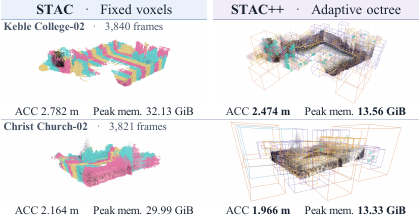}
    \caption{\textbf{STAC vs.\ STAC++ on two Oxford sequences.}
    STAC++ achieves lower reconstruction error (ACC) and uses less
    peak GPU memory on both sequences.}
    \label{fig:octree_memory}
\end{figure}

%% file: sec/2_related.tex
\section{Related Work}
\label{sec:related}

\subsection{Offline Feed-Forward 3D Reconstruction}
\label{sec:related_offline}

DUSt3R~\cite{dust3r} predicts pairwise pointmaps with global alignment; MASt3R~\cite{mast3r} adds learned matching features.
VGGT, Fast3R, and MUSt3R~\cite{vggt,fast3r,must3r} extend prediction to multiple views, while FLARE~\cite{flare} jointly recovers geometry, appearance, and cameras.
$\pi^3$~\cite{pi3} removes reference-view dependence through permutation equivariance, and VGGT-$\Omega$~\cite{vggtomega} scales geometric learning with efficient register attention.
Dense cross-view interaction nevertheless remains costly as the image collection grows.

For scalability, MERG3R~\cite{merg3r} partitions and aligns image subsets, while LiteVGGT~\cite{litevggt} merges geometrically redundant tokens and reuses merge decisions across layers.
VGG-T$^3$, ZipMap, and Scal3R~\cite{vggt3,zipmap,scal3r} compress context into test-time-trained states, using fast weights or synchronized adaptive memory to retain information across image chunks; Test3R~\cite{test3r} adapts reconstruction models at test time.
These approaches address offline scalability or adaptation rather than adaptive causal KV storage.
Octree-GS~\cite{octreegs} provides hierarchical spatial detail for Gaussian rendering; our hierarchy instead compresses transformer KV memory for subsequent streaming inference.

\subsection{Online Feed-Forward 3D Reconstruction}
\label{sec:related_online}

Spann3R~\cite{spann3r} queries external spatial memory to predict globally aligned pointmaps, CUT3R~\cite{cut3r} maintains a recurrent scene state, and Point3R~\cite{point3r} introduces explicit spatial pointer memory.
TTT3R~\cite{ttt3r} uses confidence-guided updates, while FILT3R~\cite{filt3r} applies uncertainty-adaptive Kalman-style state updates without retraining.
SLAM3R~\cite{slam3r} learns local reconstruction and registration; MASt3R-SLAM and multi-agent MAGiSt3R~\cite{mast3rslam,magist3r} additionally use geometric optimization.
LASER~\cite{laser} enables training-free streaming through layer-wise depth-scale alignment across windows.
STream3R and StreamVGGT~\cite{stream3r,streamvggt} use causal KV caching; LongStream~\cite{longstream} adds cache-aware training and refresh, while LingBot-Map~\cite{lingbotmap} separates anchor, local, and trajectory context for stable long-sequence reconstruction.

Growing caches motivate selective retention.
In language models, H$_2$O~\cite{h2o}, StreamingLLM~\cite{streamingllm}, and SnapKV~\cite{snapkv} retain attention-heavy tokens, attention sinks, or head-specific evidence.
For geometry, InfiniteVGGT~\cite{infinitevggt} uses diversity-aware pruning; OVGGT~\cite{ovggt} protects anchors during residual-guided selection; RegVGGT~\cite{regvggt} regulates incoming tokens.
GHOST~\cite{ghost} performs geometry-guided eviction, whereas RetrieveVGGT~\cite{retrievevggt} retrieves context through query--key similarity and poses.
Selection reduces cache growth but can discard useful geometric evidence; spatial aggregation offers a complementary way to retain information from repeated observations.
Slot3R~\cite{slot3r} avoids prematurely fusing complementary views by preserving multiple states per spatial address, and SURE-Map~\cite{suremap} uses cross-view uncertainty for local correction and periodic keyframe-window inference for consistent geometry.

Most closely related to our work, STAC~\cite{stac} combines temporal token retention with spatial KV fusion on a fixed-resolution voxel grid, enabling historical reuse without retraining the backbone.
We extend this design with adaptive spatial resolution and response-guided KV merging to control memory growth, and selective suffix refinement to improve earlier predictions using retained evidence.

%% file: sec/3_preliminaries.tex
\section{Preliminaries}
\label{sec:preliminaries}

\subsection{Causal Attention}
\label{sec:causal_geometry}

Causal attention enables streaming reconstruction by restricting each frame to information from the current and preceding frames, as adopted in STream3R~\cite{stream3r} and StreamVGGT~\cite{streamvggt}.
Let $F_t$ denote the encoded features of frame $t$ and $G_t$ its geometric predictions.
For a sequence of $T$ frames, this process can be summarized as
\begin{equation}
 \{G_t\}_{t=1}^{T}
 =\operatorname{Decoder}\!\left(
 \operatorname{TemporalSelfAttn}\!\left(\{F_t\}_{t=1}^{T}\right)
 \right),
 \label{eq:causal_attention}
\end{equation}
where $\operatorname{TemporalSelfAttn}$ applies a causal mask so that frame $t$ attends only to frames $1,\ldots,t$, and $\operatorname{Decoder}$ maps the attended features to per-frame geometric predictions.
During streaming inference, historical keys and values are cached and reused without reprocessing preceding frames.
Retaining the full KV history, however, makes cache storage grow linearly with sequence length.

\subsection{Spatio-Temporal KV Caching}
\label{sec:stac_cache}

STAC~\cite{stac} maintains complementary temporal and spatial KV caches.
The temporal cache $\mathcal M_t^{\mathrm{temp}}$ retains first-frame references, recent frames and anchors.
The spatial cache organizes temporally evicted patch tokens into voxels of side length $\delta$ according to their predicted 3D positions.
Retrieval from this cache provides the spatial context $\mathcal M_t^{\mathrm{spat}}$.

\textbf{Spatial cache.}
Each voxel $u$ stores up to $P$ KV representatives $\mathcal G_u$ and $B$ buffered tokens $\mathcal E_u$, consolidating tokens through aggregation when capacity is reached.
Representatives maintain fusion weights $W_i$, importance scores $S_i$, and represented-token counts $C_i$, while buffered tokens retain importance scores and contribute only through aggregation.
At frame $t$, retrieval searches the occupied voxels $\mathcal N_t$ within neighborhoods extending two voxel steps ($2\delta$) in both directions along each axis around voxels containing the preceding frame's predicted 3D points.
The spatial context is formed by selecting at most $R$ representatives per attention head with the largest fusion weights:
\begin{equation}
\mathcal M_t^{\mathrm{spat}}
=
\operatorname{Top}_{R}^{\,W}\!\left(
\bigcup_{u\in\mathcal N_t}\mathcal G_u
\right).
\label{eq:spatial_retrieval}
\end{equation}

The retrieved spatial states, temporal cache, and current-frame KV $\mathcal M_t^{\mathrm{cur}}$ form the attention context
$\mathcal C_t=\mathcal M_t^{\mathrm{temp}}\cup\mathcal M_t^{\mathrm{spat}}\cup\mathcal M_t^{\mathrm{cur}}$.
This work focuses on the spatial cache.

%% file: sec/4_method.tex
\section{Method}
\label{sec:method}
\input{fig/pipeline}

We propose \textbf{STAC++}, a \textbf{plug-and-play}, training-free framework for controlling spatial KV memory growth and reusing historical evidence in streaming 3D reconstruction.
STAC++ combines three modules (Fig.~\ref{fig:pipeline}).
\textbf{Adaptive Spatial KV Cache (ASKC)} organizes historical states at multiple spatial resolutions.
\textbf{Response-Aware Cache Compression (RACC)} determines which states can be merged by measuring their effect on attention outputs.
\textbf{Temporal Context Refinement (TCR)} revisits historical frames after a delay, reusing cached intermediate features and recomputing a short backbone suffix to incorporate later observations.
All three modules operate during inference and preserve the pretrained backbone weights.

\subsection{Adaptive Spatial KV Cache (ASKC)}
\label{sec:octree}

In unbounded scenes, fixed-resolution spatial caches grow with scene coverage and retain redundancy across neighboring voxels.
Uniform coarsening saves storage but may discard locally distinct geometric evidence.
ASKC instead organizes spatial KV states in an expanding octree, selectively merging neighboring fine-grained cells into coarser parents independently across layers while retaining finer representations where detailed evidence remains necessary.

\textbf{Layer-wise modeling.}
For a backbone with $L$ global-attention layers, layer $\ell$ maintains octree $\mathcal T_t^{(\ell)}$ at time $t$, with finest cell side length $\delta$ (Sec.~\ref{sec:stac_cache}).
Layers share the same subdivision rule but merge independently, allowing each layer to use a different spatial resolution for the same scene region.
All $H$ heads in a layer share cells but store separate KV states.
We denote the layer's token inputs by $X_t^{(\ell)}$.
Temporally evicted patch tokens enter cells containing their predicted 3D positions, including parents formed by earlier merges.
We repeatedly double the shared root's side length toward an out-of-bounds point until it is enclosed, retaining the old root as a child at each expansion step.

\textbf{Within-cell storage.}
Each cell $u$ in layer $\ell$ and head $h$ stores up to $P$ representatives $\mathcal G_{u,\ell,h}$ and $B$ buffered tokens $\mathcal E_{u,\ell,h}$.
Following STAC's online aggregation~\cite{stac}, we omit cell, layer, and head indices and write each representative as $m_i=(k_i,v_i)$, with key and value vectors $k_i$ and $v_i$.
When a new representative needs a slot but all $P$ slots are occupied, we fuse $i$ with smallest $W_i$ into the remaining representative $j$ with highest key cosine similarity:
\begin{equation}
m_j\leftarrow
\frac{W_jm_j+\omega_{ij}m_i}{W_j+\omega_{ij}},
\label{eq:fusion}
\end{equation}
where $\omega_{ij}=W_i\exp(\cos(k_i,k_j)-1)$.
We update $W_j\leftarrow W_j+\omega_{ij}$,
$C_j\leftarrow C_j+C_i$, and $S_j\leftarrow S_j+S_i$.
An incoming token sufficiently similar to a representative uses the same update with
$\omega_{ij}=\exp(\cos(k_i,k_j))$ and $C_i=1$.
Sufficiently dissimilar tokens passing the importance threshold enter the buffer.
When full, its highest-scoring token seeds aggregation of all entries into a new representative using exponential cosine-similarity weights.

\textbf{Spatial retrieval.}
Let $\Omega_t$ be the lookup neighborhoods of the preceding frame's predicted 3D points, expanded by $2\delta$ (Sec.~\ref{sec:stac_cache}), and $\mathcal N_t^{(\ell)}$ all active cells whose extents intersect $\Omega_t$, including merged parents.
For head $h$, Eq.~\ref{eq:spatial_retrieval} with $\mathcal N_t^{(\ell)}$ and $\mathcal G_{u,\ell,h}$ replacing $\mathcal N_t$ and $\mathcal G_u$, respectively, yields $\mathcal M_{t,\ell,h}^{\mathrm{spat}}$ under the same per-head budget $R$.

\textbf{Hierarchical coarsening.}
For each layer, an inactive parent is structurally eligible for merging if it has one to eight active direct children and no KV in its grandchildren or deeper descendants.
Their consolidation into the parent is a \emph{parent proposal} requiring RACC acceptance (Sec.~\ref{sec:response}).

\subsection{Response-Aware Cache Compression (RACC)}
\label{sec:response}

Merging spatially adjacent states saves memory but can alter the geometric evidence available to the backbone.
A straightforward approach is to explicitly compute the attention responses before and after a candidate merge $p$ and take their difference:
\begin{equation}
\Delta o_{p,\ell,h}
={}\operatorname{Att}(\widehat q_{\ell,h},\mathcal C_{p,\ell,h})-\operatorname{Att}(\widehat q_{\ell,h},\mathcal C_{\ell,h}),
\label{eq:response_change}
\end{equation}
where $\operatorname{Att}(\cdot)$ denotes attention, $\widehat q_{\ell,h}$ is the shared query, and $\mathcal C_{\ell,h}$ and $\mathcal C_{p,\ell,h}$ are the contexts before and after the merge.
The corresponding responses are denoted by $o_{\ell,h}$ and $o_{p,\ell,h}$, respectively.
However, directly evaluating this change for every candidate merge requires repeating query--key scoring, softmax normalization, and value aggregation over the full context, making evaluation costly.
We therefore sample a small set of queries and compute response changes incrementally, reusing the reference response and normalizer while updating only the contributions of KV entries affected by each merge.
RACC balances response preservation against storage savings.

\textbf{Merge scheduling.}
For merge keyframes spaced $M$ frames apart, we retain $\Omega_{t-M}$ and full inputs $X_{t-M}^{(\ell)}$ to each global-attention layer.
After keyframe $t$ completes inference, we sample queries from these inputs at all special-token positions and a uniform $4\times4$ patch grid to evaluate eligible parents in $\Omega_{t-M}$.
This delay allows intervening frames to use the finer spatial representation before coarsening.
Pending keyframes use the available context at sequence end.

\textbf{Parent proposal construction.}
At each layer $\ell$, consider an eligible parent $U$ in $\Omega_{t-M}$ with child positions $u_i$ ($i=1,\ldots,8$; empty branches give empty sets), omitting $t-M$ below.
Head $h$ retains up to $B$ buffered entries with highest importance from $\bigcup_{i=1}^{8}\mathcal E_{u_i,\ell,h}$ in $\mathcal E_{U,\ell,h}$.
Child representatives form $\mathcal G_{U,\ell,h}$ unchanged when $\sum_{i=1}^{8}|\mathcal G_{u_i,\ell,h}|\leq P$; otherwise, we repeatedly apply the representative re-merging rule in Sec.~\ref{sec:octree} using Eq.~\ref{eq:fusion} until $P$ remain.
The pairs $p_{U,\ell,h}=(\mathcal G_{U,\ell,h},\mathcal E_{U,\ell,h})$ form proposal $p$, saving
\begin{equation}
\begin{aligned}
\Delta N_{p,\ell,h}
={}&
\sum_{i=1}^{8}
\left(
|\mathcal G_{u_i,\ell,h}|
+
|\mathcal E_{u_i,\ell,h}|
\right)\\
&-
|\mathcal G_{U,\ell,h}|
-
|\mathcal E_{U,\ell,h}|.
\end{aligned}
\label{eq:proposal_savings}
\end{equation}
Summing over heads gives
$\Delta N_{p,\ell}=\sum_{h=1}^{H}\Delta N_{p,\ell,h}$.

\textbf{Response change evaluation.}
For sampled $\widehat q_{\ell,h}$, compute $o_{\ell,h}=\operatorname{Att}(\widehat q_{\ell,h},\mathcal C_{\ell,h})$ over $\mathcal C_{\ell,h}=\mathcal M_{\ell,h}^{\mathrm{temp}}\cup\mathcal M_{\ell,h}^{\mathrm{spat}}\cup\mathcal M_{\ell,h}^{\mathrm{cur}}$ (time omitted).
We retain $\Lambda_{\ell,h}=\log\sum_{i\in\mathcal C_{\ell,h}}\exp(z_i)$, where $z_i=\widehat q_{\ell,h}^{\top}k_i/\sqrt d+\log C_i$, suppressing layer/head indices on tokens; $d$ is the key dimension and $C_i=1$ for unmerged tokens.
Proposal $p$ replaces retrieved child representatives $\mathcal R_p$ with distinct parent representatives $\mathcal A_p\subseteq\mathcal G_{U,\ell,h}$ containing their contributions, giving $\mathcal C_{p,\ell,h}=(\mathcal C_{\ell,h}\setminus\mathcal R_p)\cup\mathcal A_p$.
The remaining context and retrieval ranking stay fixed.
For $i\in\mathcal R_p\cup\mathcal A_p$, set $\alpha_i=\exp(z_i-\Lambda_{\ell,h})$ using the reference normalizer.
Equation~\ref{eq:response_change} becomes
\begin{equation}
\Delta o_{p,\ell,h}
=
\frac{
\displaystyle\sum_{i\in\mathcal A_p}\alpha_i(v_i-o_{\ell,h})
-\displaystyle\sum_{i\in\mathcal R_p}\alpha_i(v_i-o_{\ell,h})
}{
1+\displaystyle\sum_{i\in\mathcal A_p}\alpha_i
-\displaystyle\sum_{i\in\mathcal R_p}\alpha_i
}.
\label{eq:delta}
\end{equation}
Equation~\ref{eq:delta} is exact for the fixed query and specified substitution~\cite{online_softmax,flashattention}; only $\mathcal R_p$ and $\mathcal A_p$ require new scores and value accumulations.
Unchanged contributions are reused through $o_{\ell,h}$ and $\Lambda_{\ell,h}$.
Stacking reference responses and changes over sampled queries yields $O_{\ell,h}$ and $\Delta O_{p,\ell,h}$.

\textbf{Merge acceptance.}
We average normalized response changes across heads:
\begin{equation}
E_{p,\ell}
=
\frac{1}{H}\sum_{h=1}^{H}
\frac{\lVert\Delta O_{p,\ell,h}\rVert_F}
{\max(\lVert O_{\ell,h}\rVert_F,\epsilon_m)},
\label{eq:mergeerror}
\end{equation}
with $\epsilon_m=10^{-8}$ for stability. We accept $p$ when
\begin{equation}
E_{p,\ell}\leq
\tau_m+\lambda\frac{\Delta N_{p,\ell}}{H(P+B)}.
\label{eq:budget}
\end{equation}
Here $\tau_m$ sets the baseline tolerance and $\lambda$ weights storage savings.
Each layer makes an independent decision shared across all heads; acceptance replaces child storage cells with their parent.

\begin{table*}[!t]
\centering
\caption{3D reconstruction on 7-Scenes, Neural RGB-D, and ETH3D. Across four online backbones, STAC++ uses less memory than STAC while maintaining competitive reconstruction quality. Memory denotes mean peak GPU usage in GiB.
}
\label{tab:standard_reconstruction}
\begingroup
\fontsize{8}{9.2}\selectfont
\setlength{\tabcolsep}{9.2pt}
\renewcommand{\arraystretch}{1.1}
\edef\STACmetricwidth{\the\dimexpr(\textwidth-0.22\textwidth-0.065\textwidth-22\tabcolsep)/10\relax}
\edef\STACheaderdrop{\the\dimexpr-\arraystretch\dimexpr\baselineskip/2\relax\relax}
\begin{tabular}{@{}l|c|ccc|ccc|ccc|c@{}}
\toprule
\raisebox{\STACheaderdrop}[0pt][0pt]{\textbf{Method}} & \raisebox{\STACheaderdrop}[0pt][0pt]{\makebox[0.065\textwidth][c]{\textbf{Type}}} & \multicolumn{3}{c|}{\textbf{7-Scenes}} & \multicolumn{3}{c|}{\textbf{Neural RGB-D}} & \multicolumn{3}{c|}{\textbf{ETH3D}} & \makebox[\STACmetricwidth][c]{\textbf{Mem} $\boldsymbol{\downarrow}$} \\
 & & \makebox[\STACmetricwidth][c]{\textbf{Acc} $\downarrow$} & \makebox[\STACmetricwidth][c]{\textbf{Comp} $\downarrow$} & \makebox[\STACmetricwidth][c]{\textbf{F1} $\uparrow$} & \makebox[\STACmetricwidth][c]{\textbf{Acc} $\downarrow$} & \makebox[\STACmetricwidth][c]{\textbf{Comp} $\downarrow$} & \makebox[\STACmetricwidth][c]{\textbf{F1} $\uparrow$} & \makebox[\STACmetricwidth][c]{\textbf{Acc} $\downarrow$} & \makebox[\STACmetricwidth][c]{\textbf{Comp} $\downarrow$} & \makebox[\STACmetricwidth][c]{\textbf{F1} $\uparrow$} & (GiB) \\
\midrule
ViPE~\cite{vipe} & Optim & 0.061 & 0.032 & 85.56 & 0.183 & 0.105 & 61.05 & 2.514 & 3.493 & 12.47 & 11.38 \\
MegaSaM~\cite{megasam} & Optim & 0.027 & 0.026 & 94.25 & 0.047 & 0.018 & 93.81 & 1.910 & 1.981 & 19.84 & 10.70 \\
\midrule
VGGT~\cite{vggt} & Offline & 0.022 & 0.022 & 96.46 & 0.017 & 0.012 & 99.04 & 0.199 & 0.128 & 86.91 & 16.88 \\
VGG-T$^3$~\cite{vggt3} & Offline & 0.046 & 0.029 & 92.06 & 0.059 & 0.025 & 86.15 & 0.441 & 0.292 & 70.47 & 14.99 \\
\midrule
CUT3R~\cite{cut3r} & Online & 0.093 & 0.038 & 81.09 & 0.210 & 0.095 & 62.44 & 0.933 & 0.730 & 58.57 & 10.34 \\
TTT3R~\cite{ttt3r} & Online & 0.035 & 0.026 & 92.99 & 0.101 & 0.031 & 83.84 & 0.607 & 0.597 & 63.20 & 6.02 \\
ZipMap~\cite{zipmap} & Online & 0.114 & 0.054 & 76.14 & 0.368 & 0.271 & 47.50 & 0.807 & 0.570 & 60.83 & 17.15 \\
\midrule
STream3R~\cite{stream3r}\textsuperscript{a} & Online & \textbf{0.043} & \underline{0.021} & \textbf{95.52} & 0.033 & 0.010 & 97.30 & 0.553 & 0.288 & \underline{70.33} & 27.18 \\
STream3R-W & Online & 0.082 & 0.025 & 89.63 & 0.085 & 0.017 & 88.86 & 0.556 & 0.293 & 70.24 & \underline{11.67} \\
STream3R-STAC & Online & 0.048 & \textbf{0.020} & \underline{95.38} & \underline{0.066} & \textbf{0.013} & \textbf{94.56} & \textbf{0.492} & \underline{0.242} & \textbf{70.59} & 14.28 \\
STream3R-STAC++ & Online & \underline{0.047} & \textbf{0.020} & 95.32 & \textbf{0.063} & \underline{0.014} & \underline{94.14} & \underline{0.543} & \textbf{0.234} & 69.74 & \textbf{10.84} \\
\midrule
StreamVGGT~\cite{streamvggt}\textsuperscript{b} & Online & \textbf{0.046} & \textbf{0.025} & \textbf{90.74} & 0.132 & 0.064 & 70.44 & 0.674 & 0.504 & \underline{56.96} & 23.41 \\
StreamVGGT-W & Online & 0.071 & 0.031 & 85.59 & 0.165 & 0.075 & 62.85 & 0.676 & 0.503 & 56.87 & \underline{11.41} \\
StreamVGGT-STAC & Online & 0.056 & \underline{0.026} & 89.21 & \underline{0.141} & \underline{0.056} & \underline{68.12} & \underline{0.604} & \underline{0.471} & \textbf{59.59} & 15.17 \\
StreamVGGT-STAC++ & Online & \underline{0.053} & 0.027 & \underline{89.76} & \textbf{0.126} & \textbf{0.046} & \textbf{73.16} & \textbf{0.596} & \textbf{0.458} & 56.64 & \textbf{11.11} \\
\midrule
InfiniteVGGT~\cite{infinitevggt} & Online & \textbf{0.048} & \textbf{0.025} & \textbf{90.27} & \underline{0.126} & \underline{0.059} & 69.40 & 0.637 & 0.474 & 57.14 & \underline{13.57} \\
InfiniteVGGT-STAC & Online & 0.051 & \underline{0.027} & 89.88 & 0.129 & 0.060 & \underline{69.42} & \underline{0.627} & \underline{0.463} & \textbf{58.23} & 15.43 \\
InfiniteVGGT-STAC++ & Online & \underline{0.049} & 0.028 & \underline{90.09} & \textbf{0.121} & \textbf{0.050} & \textbf{72.18} & \textbf{0.610} & \textbf{0.442} & \underline{57.35} & \textbf{12.75} \\
\midrule
LingBot-Map~\cite{lingbotmap} & Online & \underline{0.029} & \textbf{0.020} & 94.96 & 0.057 & \underline{0.023} & 89.08 & \underline{0.487} & 0.514 & \underline{69.97} & \textbf{13.35} \\
LingBot-Map-STAC & Online & \textbf{0.028} & \underline{0.021} & \textbf{95.13} & \underline{0.055} & \textbf{0.021} & \underline{90.93} & \underline{0.487} & \underline{0.485} & 69.53 & 17.59 \\
LingBot-Map-STAC++ & Online & \textbf{0.028} & \underline{0.021} & \underline{95.08} & \textbf{0.054} & \textbf{0.021} & \textbf{91.07} & \textbf{0.480} & \textbf{0.479} & \textbf{70.68} & \underline{15.12} \\
\bottomrule
\end{tabular}
\par\vspace{3pt}
\begin{minipage}{\linewidth}
\footnotesize\raggedright
\textsuperscript{a} STream3R encounters OOM on 4 of 9 Neural RGB-D \par
\textsuperscript{b} StreamVGGT encounters OOM on 1 of 9 Neural RGB-D
\end{minipage}
\endgroup
\end{table*}

\subsection{Temporal Context Refinement (TCR)}
\label{sec:refine}

During causal streaming inference, each frame is reconstructed using only the observations available when it arrives.
Later frames may provide complementary viewpoints and additional geometric evidence for previously observed regions, enriching the spatial cache.
However, this updated memory benefits subsequent inference without automatically revising earlier predictions, which remain tied to their original, more limited context.
\emph{The most direct approach is to reprocess historical frames through the entire backbone using the updated context.}
However, this repeats substantial feature computation.
We therefore introduce TCR to refine historical predictions after a fixed delay using the updated temporal and spatial context.
TCR retains intermediate features during initial inference and recomputes only a short backbone suffix, allowing earlier frames to reuse later observations while limiting the additional computation.

\textbf{Refinement scheduling.}
During initial inference, we retain each frame's decoder features from layers 5 and 12, the intermediate input needed for suffix recomputation, and its spatial lookup region.
After a delay of $D$ frames, we revisit the historical frame once using the updated context.
The refinement delay is independent of the merge interval, and merge keyframes share their retained inputs with RACC.

\textbf{Refinement process.}
At time $t$, we refine frame $t-D$ by resuming from its retained intermediate features.
At each recomputed attention layer, queries from this revisited frame attend to the current temporal cache, spatial states retrieved within its saved lookup region, and keys and values recomputed from the same frame's features at that layer.
Let $X$ denote the retained intermediate input and $\mathcal C$ the updated context.
Recomputing layers 18--24 gives
\begin{equation}
\bigl(\widetilde F^{(18)},\widetilde F^{(24)}\bigr)
=\operatorname{Suffix}_{18:24}(X,\mathcal C),
\label{eq:suffix}
\end{equation}
where $\operatorname{Suffix}_{18:24}$ denotes recomputation of backbone layers 18--24 from the retained input $X$ using the updated context $\mathcal C$, returning the features at layers 18 and 24.
Here $F^{(\ell)}$ denotes the feature at layer $\ell$, tildes mark refinement, and the frame index is omitted.
The decoder combines the cached features from layers 5 and 12 with the updated features from layers 18 and 24:
\begin{equation}
\widetilde G=\operatorname{Decoder}\!\left(
F^{(5)},F^{(12)},\widetilde F^{(18)},\widetilde F^{(24)}
\right),
\label{eq:refined_prediction}
\end{equation}
where $\widetilde G$ denotes the refined geometric predictions produced by the native dense and camera heads.

%% file: fig/pipeline.tex
\begin{figure*}[!t]
    \centering
    \includegraphics[width=\textwidth]{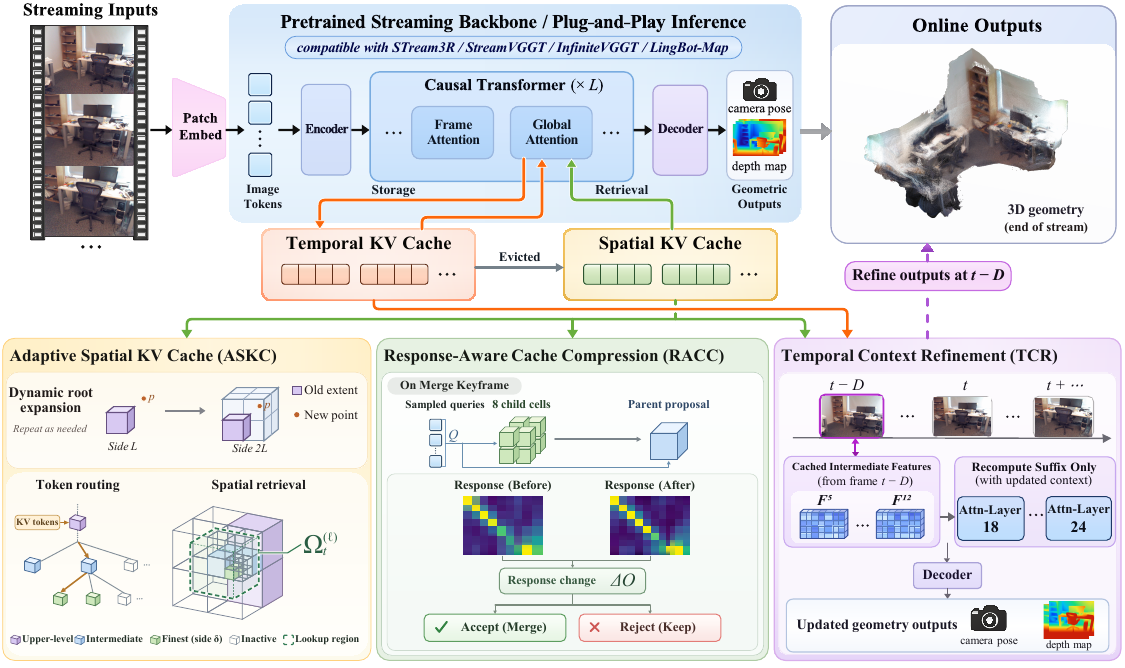}
    \caption{\textbf{Overview of STAC++.} A plug-and-play, training-free framework for streaming 3D reconstruction.
    ASKC organizes spatial KV states in an adaptive octree; RACC guides cache compression using attention-output changes.
    TCR reuses cached features and recomputes a short backbone suffix to refine historical predictions with later observations.}
    \label{fig:pipeline}
\end{figure*}

%% file: sec/5_experiments.tex
\begin{figure*}[!t]
    \centering
    \includegraphics[pagebox=cropbox,width=\textwidth]{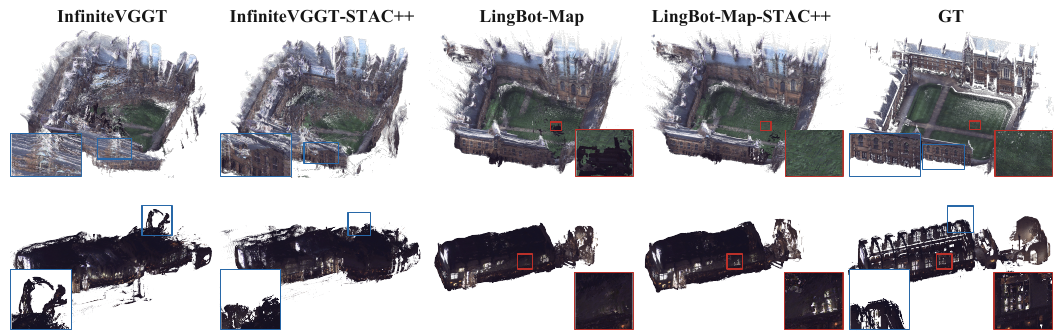}
    \caption{Qualitative comparison of large-scene reconstructions.
    Blue insets compare InfiniteVGGT with its STAC++ variant,
    and red insets compare LingBot-Map with its STAC++ variant.
    The rightmost column shows the laser-scan ground truth.}
    \label{fig:qualitative_results}
\end{figure*}

\begin{table}[t]
\centering
\caption{Quantitative results on Long3D and Oxford Spires. STAC++ completes all evaluated sequences with InfiniteVGGT and LingBot-Map, while their STAC variants run out of memory. Memory is measured in GiB and inference time in ms/frame.}
\label{tab:large_reconstruction}
\begingroup
\providecommand{\STACna}{\textcolor{black!40}{\textit{NA}}}
\providecommand{\STACoom}{\textcolor{black!40}{\textit{OOM}}}
\footnotesize
\setlength{\tabcolsep}{1pt}
\renewcommand{\arraystretch}{1.05}
\begin{tabular*}{\linewidth}{@{\extracolsep{\fill}}lccccc@{}}
\toprule
\textbf{Method} & \textbf{Acc} $\downarrow$ & \textbf{Comp} $\downarrow$ & \textbf{F1} $\uparrow$ & \textbf{Mem} $\downarrow$ & \textbf{Inference} $\downarrow$ \\
\midrule
\multicolumn{6}{@{}l}{\textbf{Long3D}} \\
\midrule
STream3R~\cite{stream3r} & \STACoom & \STACoom & \STACoom & \STACoom & \STACoom \\
STream3R-W & \textbf{2.815} & \textbf{0.768} & \textbf{26.06} & \underline{13.32} & \underline{224.26} \\
STream3R-STAC & 3.210 & 0.917 & 22.57 & 15.69 & 116.84 \\
STream3R-STAC++ & \underline{3.192} & \underline{0.858} & \underline{22.63} & \textbf{10.75} & \textbf{190.52} \\
\cmidrule[0.3pt](lr){1-6}
StreamVGGT~\cite{streamvggt} & \STACoom & \STACoom & \STACoom & \STACoom & \STACoom \\
StreamVGGT-W & \underline{2.717} & \textbf{0.648} & \underline{23.76} & \underline{13.27} & \underline{285.96} \\
StreamVGGT-STAC\textsuperscript{d} & 2.691 & 0.533 & 25.82 & 29.45 & 118.49 \\
StreamVGGT-STAC++ & \textbf{2.713} & \underline{1.003} & \textbf{24.30} & \textbf{11.39} & \textbf{205.04} \\
\cmidrule[0.3pt](lr){1-6}
InfiniteVGGT~\cite{infinitevggt} & \textbf{3.136} & \textbf{0.767} & \textbf{22.08} & \underline{15.93} & \textbf{419.29} \\
InfiniteVGGT-STAC\textsuperscript{e} & 3.276 & 1.106 & 23.87 & 21.19 & 386.82 \\
InfiniteVGGT-STAC++ & \underline{3.242} & \underline{1.559} & \underline{20.67} & \textbf{15.72} & \underline{472.00} \\
\cmidrule[0.3pt](lr){1-6}
LingBot-Map~\cite{lingbotmap} & \textbf{2.007} & \underline{0.968} & \textbf{29.89} & \textbf{11.92} & \underline{252.26} \\
LingBot-Map-STAC & \STACoom & \STACoom & \STACoom & \STACoom & \STACoom \\
LingBot-Map-STAC++ & \underline{2.354} & \textbf{0.948} & \underline{29.39} & \underline{16.48} & \textbf{239.67} \\
\midrule
\multicolumn{6}{@{}l}{\textbf{Oxford Spires}} \\
\midrule
InfiniteVGGT~\cite{infinitevggt} & \underline{18.436} & \underline{5.907} & \underline{2.00} & \textbf{15.54} & \textbf{464.58} \\
InfiniteVGGT-STAC & \STACoom & \STACoom & \STACoom & \STACoom & \STACoom \\
InfiniteVGGT-STAC++ & \textbf{11.594} & \textbf{1.686} & \textbf{4.96} & \underline{17.40} & \underline{628.46} \\
\cmidrule[0.3pt](lr){1-6}
LingBot-Map~\cite{lingbotmap} & \underline{2.951} & \underline{1.730} & \textbf{19.59} & \textbf{15.43} & \textbf{154.54} \\
LingBot-Map-STAC\textsuperscript{c} & 3.489 & 1.665 & 14.27 & 35.47 & 257.88 \\
LingBot-Map-STAC++ & \textbf{2.783} & \textbf{1.466} & \underline{19.28} & \underline{15.65} & \underline{204.08} \\
\bottomrule
\end{tabular*}
\par\vspace{2pt}
\begin{minipage}{\linewidth}
\footnotesize\raggedright
\textsuperscript{c} LingBot-Map-STAC encounters OOM on 3/9 Oxford Spires \par
\textsuperscript{d} StreamVGGT-STAC encounters OOM on 1/5 Long3D \par
\textsuperscript{e} InfiniteVGGT-STAC encounters OOM on 1/5 Long3D
\end{minipage}
\endgroup
\end{table}

\section{Experiments}
\label{sec:experiments}

\begin{table*}[!t]
\centering
\caption{Camera pose estimation on Sintel, TUM RGB-D, and ETH3D. Across four online backbones, STAC++ generally lowers ATE while using less memory than STAC. We report ATE (m), angular AUC (\%), and mean peak GPU memory (GiB).
}
\label{tab:standard_camera}
\begingroup
\fontsize{8}{9.2}\selectfont
\setlength{\tabcolsep}{2pt}
\renewcommand{\arraystretch}{1.1}
\edef\STACmetricwidth{\the\dimexpr(\textwidth-0.22\textwidth-0.065\textwidth-22\tabcolsep)/10\relax}
\edef\STACheaderdrop{\the\dimexpr-\arraystretch\dimexpr\baselineskip/2\relax\relax}
\begin{tabular}{@{}l|c|ccc|ccc|ccc|c@{}}
\toprule
\raisebox{\STACheaderdrop}[0pt][0pt]{\textbf{Method}} & \raisebox{\STACheaderdrop}[0pt][0pt]{\makebox[0.065\textwidth][c]{\textbf{Type}}} & \multicolumn{3}{c|}{\textbf{Sintel}} & \multicolumn{3}{c|}{\textbf{TUM}} & \multicolumn{3}{c|}{\textbf{ETH3D}} & \makebox[\STACmetricwidth][c]{\textbf{Mem} $\downarrow$} \\
 & & \makebox[\STACmetricwidth][c]{\textbf{ATE} $\downarrow$} & \textbf{AUC@3}\,$\uparrow$ & \textbf{AUC@30}\,$\uparrow$ & \makebox[\STACmetricwidth][c]{\textbf{ATE} $\downarrow$} & \textbf{AUC@3}\,$\uparrow$ & \textbf{AUC@30}\,$\uparrow$ & \makebox[\STACmetricwidth][c]{\textbf{ATE} $\downarrow$} & \textbf{AUC@3}\,$\uparrow$ & \textbf{AUC@30}\,$\uparrow$ & (GiB) \\
\midrule
STream3R~\cite{stream3r} & Online & 0.214 & \textbf{7.65} & \textbf{53.02} & \textbf{0.025} & \textbf{7.10} & \textbf{48.62} & 1.906 & 14.26 & 62.45 & 13.49 \\
STream3R-W & Online & 0.214 & \underline{7.41} & 52.76 & 0.032 & 3.93 & 41.80 & 1.855 & 14.08 & 61.80 & \underline{10.61} \\
STream3R-STAC & Online & \underline{0.204} & 6.12 & \underline{53.01} & 0.032 & \underline{6.06} & 44.70 & \textbf{1.433} & \underline{16.78} & \textbf{67.25} & 12.33 \\
STream3R-STAC++ & Online & \textbf{0.201} & 6.27 & 51.75 & \underline{0.031} & 4.78 & \underline{44.89} & \underline{1.551} & \textbf{17.00} & \underline{66.80} & \textbf{9.07} \\
\midrule
StreamVGGT~\cite{streamvggt} & Online & 0.233 & 6.00 & 47.94 & \underline{0.026} & 8.12 & 53.48 & 1.898 & 13.59 & 58.83 & 11.83 \\
StreamVGGT-W & Online & 0.233 & 6.00 & 47.94 & 0.029 & 7.76 & 50.71 & 1.900 & 13.51 & 58.58 & \underline{10.35} \\
StreamVGGT-STAC & Online & \underline{0.226} & \textbf{10.03} & \textbf{52.63} & \underline{0.026} & \textbf{12.32} & \textbf{58.44} & \underline{1.649} & \textbf{16.46} & \underline{62.25} & 13.22 \\
StreamVGGT-STAC++ & Online & \textbf{0.214} & \underline{8.26} & \underline{49.42} & \textbf{0.025} & \underline{10.38} & \underline{56.11} & \textbf{1.607} & \underline{15.85} & \textbf{64.20} & \textbf{9.65} \\
\midrule
InfiniteVGGT~\cite{infinitevggt} & Online & \underline{0.236} & \underline{6.36} & 47.21 & \underline{0.027} & \textbf{8.25} & 52.60 & \underline{1.742} & \underline{12.87} & 59.63 & \textbf{12.15} \\
InfiniteVGGT-STAC & Online & 0.238 & 6.33 & \underline{47.25} & \underline{0.027} & \underline{8.13} & \textbf{52.74} & 1.755 & 12.61 & \underline{59.67} & 16.02 \\
InfiniteVGGT-STAC++ & Online & \textbf{0.229} & \textbf{7.40} & \textbf{48.00} & \textbf{0.026} & 7.66 & \underline{52.61} & \textbf{1.724} & \textbf{13.70} & \textbf{61.79} & \underline{12.63} \\
\midrule
LingBot-Map~\cite{lingbotmap} & Online & \underline{0.126} & \underline{3.56} & \underline{50.44} & 0.021 & 11.72 & 59.20 & 0.712 & 35.99 & 75.49 & \textbf{14.26} \\
LingBot-Map-STAC & Online & 0.130 & 3.47 & \textbf{50.97} & \underline{0.019} & \underline{13.43} & \underline{62.39} & \underline{0.685} & \underline{38.08} & \textbf{77.40} & 18.66 \\
LingBot-Map-STAC++ & Online & \textbf{0.121} & \textbf{3.70} & 49.68 & \textbf{0.018} & \textbf{14.41} & \textbf{63.55} & \textbf{0.623} & \textbf{38.87} & \underline{77.38} & \underline{14.56} \\
\bottomrule
\end{tabular}
\endgroup
\end{table*}

\subsection{Implementation Details}
\label{sec:implementation}

STAC++ is a plug-and-play, training-free framework.
We integrate it into four online reconstruction backbones: STream3R~\cite{stream3r}, StreamVGGT~\cite{streamvggt}, InfiniteVGGT~\cite{infinitevggt}, and LingBot-Map~\cite{lingbotmap}, keeping their pretrained weights frozen.
For comparison, we also extend STAC's spatial caching~\cite{stac} to InfiniteVGGT and LingBot-Map while preserving their native temporal memory policies.
InfiniteVGGT evicts by token budget rather than a fixed temporal window; we archive evicted patch KV in spatial memory once their 3D positions are predicted.

\textbf{Datasets and Metrics.}
We evaluate 3D reconstruction on 7-Scenes~\cite{sevenscenes}, Neural RGB-D~\cite{neural_rgbd}, and ETH3D~\cite{eth3d}, as well as on Long3D~\cite{infinitevggt} and Oxford Spires~\cite{oxford_spires}(christ-church-05 excluded) for large scenes and long streams.
For camera pose estimation, we evaluate on ETH3D, Sintel~\cite{sintel}, and TUM RGB-D~\cite{tum}.
We report accuracy (Acc) and completeness (Comp) as mean prediction-to-reference and reference-to-prediction nearest-neighbor distances (m), respectively.
F1 (\%) uses thresholds of 0.10\,m for 7-Scenes and Neural RGB-D and 0.25\,m for ETH3D, Long3D, and Oxford Spires.
For cameras, we report absolute trajectory error (ATE, m) after Sim(3) alignment and angular AUC (\%) at $3^\circ$ and $30^\circ$, using the larger relative rotation or translation-direction error per valid frame pair.

\textbf{Baselines.}
We integrate STAC++ into STream3R~\cite{stream3r}, StreamVGGT~\cite{streamvggt}, InfiniteVGGT~\cite{infinitevggt}, and LingBot-Map~\cite{lingbotmap}.
For each backbone, we compare STAC++ with native inference and STAC~\cite{stac}, and additionally include windowed STream3R and StreamVGGT baselines.

\textbf{Implementation details.}
Following STAC~\cite{stac}, each cell stores up to $P=4$ representatives and $B=8$ buffered tokens per head.
We use a finest cell size of $\delta=0.05$, merge parameters $\tau_m=\lambda=5\times10^{-4}$, suffix depth $k=8$, and default refinement delay and merge interval $D=M=8$ frames (For large scenes $D=M=16$).
STream3R-W and StreamVGGT-W use 36- and 48-frame windows, respectively, while their STAC and STAC++ variants use eight-frame windows.
InfiniteVGGT's native token budget of $1{,}200{,}000$ is reduced to $650{,}000$ for STAC and STAC++.
All LingBot-Map variants use a temporal window of 16 keyframes and retain eight scale anchors.
In this way, STAC and STAC++ share temporal memory settings within each backbone.
All experiments run on a single NVIDIA GeForce RTX 4090 with 48\,GiB of memory.

\subsection{3D Reconstruction.}
Table~\ref{tab:standard_reconstruction} compares reconstruction on 7-Scenes, Neural RGB-D, and ETH3D. Relative to windowed STream3R and StreamVGGT and native InfiniteVGGT and LingBot-Map, STAC++ improves F1 in 9 of 12 settings. All four backbones improve on Neural RGB-D, with gains of 1.99--10.31 percentage points.
At the same temporal window or token budget as STAC, STAC++ reduces mean peak GPU memory by 14.0\%--26.8\% across backbones.
This suggests that adaptive compression can reduce spatial redundancy while retaining evidence useful for reconstruction.
When integrated into LingBot-Map, STAC++ outperforms CUT3R~\cite{cut3r}, TTT3R~\cite{ttt3r}, and ZipMap~\cite{zipmap} in accuracy, completeness, and F1 across these datasets.

For large scenes and long streams, Table~\ref{tab:large_reconstruction} reports results on Long3D~\cite{infinitevggt} and Oxford Spires~\cite{oxford_spires}.
VGGT~\cite{vggt}, VGG-T$^3$~\cite{vggt3}, CUT3R~\cite{cut3r}, and ZipMap~\cite{zipmap} encounter out-of-memory (OOM) failures on these large-scene benchmarks.
With InfiniteVGGT and LingBot-Map, STAC++ completes every evaluated sequence, whereas the STAC variants encounter out-of-memory failures on multiple sequences.
This difference highlights the importance of controlling spatial memory growth as trajectories lengthen and scene coverage expands.
STAC++ also improves reconstruction quality on Oxford Spires, reducing both Acc and Comp for InfiniteVGGT and LingBot-Map.
In terms of inference time, STAC++ is faster than windowed STream3R and StreamVGGT baselines on Long3D, while its runtime relative to native InfiniteVGGT and LingBot-Map varies across datasets.
These results show that adaptive spatial caching supports more reliable long-stream inference while retaining geometric evidence, with the quality--memory trade-off varying across scenes evaluated in these benchmarks and across different pretrained backbones.

As shown in Fig.~\ref{fig:qualitative_results}, STAC++ produces high-quality large-scene reconstructions with both InfiniteVGGT and LingBot-Map.
The results preserve coherent geometry and local structural details, demonstrating that compact spatial memory can maintain visual reconstruction quality.

\begin{table}[t]
\centering
\caption{Ablation of ASKC, RACC, and TCR for LingBot-Map-STAC++ on Oxford Spires. Memory is measured in GiB and inference time in ms/frame.}
\label{tab:ablation}
\providecommand{\STACna}{\textcolor{black!40}{\textit{NA}}}
\providecommand{\STACoom}{\textcolor{black!40}{\textit{OOM}}}
\footnotesize
\setlength{\tabcolsep}{2pt}
\renewcommand{\arraystretch}{1.05}
\begin{tabular*}{\linewidth}{@{\extracolsep{\fill}}lcccc@{}}
\toprule
\textbf{Configuration} & \textbf{Acc} $\downarrow$ & \textbf{F1} $\uparrow$ & \textbf{Mem} $\downarrow$ & \textbf{Inference} $\downarrow$ \\
\midrule
w/o ASKC (STAC)\textsuperscript{c} & 3.4888 & 14.27 & 35.47 & 257.88 \\
w/o RACC & 2.9131 & 18.62 & \textbf{14.76} & \underline{172.13} \\
Random merging & 2.8929 & 19.22 & \underline{14.87} & 190.95 \\
\midrule
w/o TCR & \underline{2.7927} & 19.22 & 15.16 & \textbf{163.54} \\
TCR (2-layer) & 2.7995 & \textbf{19.52} & 15.69 & 204.31 \\
TCR (14-layer) & 2.8118 & 18.99 & 15.44 & 216.01 \\
TCR (21-layer) & 2.8031 & 19.06 & 15.44 & 239.34 \\
TCR (24-layer) & 2.8010 & 18.94 & 15.38 & 239.96 \\
\midrule
\textbf{Full(8-layer)} & \textbf{2.7828} & \underline{19.28} & 15.65 & 204.08 \\
\bottomrule
\end{tabular*}
\end{table}

\subsection{Camera Pose Estimation.}
Following prior work~\cite{stac,lingbotmap}, we evaluate camera pose estimation on Sintel~\cite{sintel}, TUM RGB-D~\cite{tum}, and ETH3D~\cite{eth3d}.
As shown in Table~\ref{tab:standard_camera}, STAC++ lowers ATE in all 12 comparisons with the corresponding native baselines, using the windowed variants for STream3R and StreamVGGT.
Compared with STAC under the same temporal window or token budget, it lowers ATE in 11 of 12 settings and reduces mean peak GPU memory by 21.2\%--27.0\% across backbones.
This suggests that a more compact spatial cache can preserve evidence useful for camera localization while improving memory efficiency.
Angular AUC gains vary across datasets and backbones, with both metrics improving for LingBot-Map on TUM RGB-D.

\subsection{Ablation Study}
\label{sec:ablation}

We conduct ablation studies on Oxford Spires to evaluate ASKC, RACC, and TCR.
The ASKC and RACC ablations examine spatial cache compression.
The TCR ablation examines temporal context refinement.

\textbf{Effect of ASKC.}
The w/o ASKC variant corresponds to STAC, which aggregates tokens only within fixed-resolution spatial cells (Table~\ref{tab:ablation}).
Hierarchical merging substantially reduces memory relative to this baseline.

\textbf{Effect of RACC.}
The w/o RACC variant accepts every eligible merge, while random merging uses the response-guided policy's per-sequence acceptance rate, with results averaged over three seeds.
Response-guided merging achieves better Acc and F1 than random merging and w/o RACC at similar memory usage (Table~\ref{tab:ablation}).

\textbf{Effect of TCR.}
In Table~\ref{tab:ablation}, refining eight layers achieves the best Acc among the TCR variants.
Shallower refinement offers little benefit, while deeper suffixes add computation without improving quality, supporting $k=8$ as the default.

%% file: sec/6_conclusion.tex
\section{Conclusion}
\label{sec:conclusion}
We presented \method, a plug-and-play, training-free framework for long-term streaming 3D reconstruction in unbounded scenes.
Adaptive Spatial KV Cache (ASKC) organizes spatial KV states in an octree with adaptive resolutions across regions and transformer layers.
Response-Aware Cache Compression (RACC) guides KV merging through attention-output changes to balance memory savings with geometric preservation.
Temporal Context Refinement (TCR) reuses cached intermediate features and recomputes a short backbone suffix to refine historical predictions with later observations.
Experiments across four pretrained backbones demonstrate reduced memory consumption and competitive reconstruction and camera pose estimation quality, with improved scalability to long sequences and large scenes.
Delayed refinement further improves reconstruction quality with limited computational overhead.